\documentclass[aps,prb,twocolumn,showpacs,amsmath,amssymb,superscriptaddress]{revtex4}
\usepackage{graphicx}
\usepackage{amsfonts}
\usepackage{amssymb}
\usepackage{Jonasmacros}
\usepackage{bm}

\begin{document}
\title{Signatures of band-like tunnelling in granular nanowires}
\author{J. Fransson}
\email{Jonas.Fransson@fysik.uu.se}
\affiliation{Department of Materials Science and Engineering, Royal Institute of Technology (KTH), SE-100 44\ \ Stockholm, Sweden}
\affiliation{Physics Department, Uppsala University, Box 530, SE-751 21\ \ Uppsala, Sweden}
\affiliation{NORDITA, Blegdamsvej 17, DK-2100\ \ Copenhagen, Denmark}
\author{J.-F. Lin}
\affiliation{Nanostructures Research Group, Department of Electrical Engineering, Arizona State University, Tempe, AZ 85287-5706, USA}
\author{L. Rotkina}
\affiliation{Beckman Institute for Advanced Science and Technology, University of Illinois at Urbana-Champaign, 405 N. Mathews Ave., Urbana, IL 61801, USA}
\author{J. P. Bird}
\affiliation{Department of Electrical Engineering, University at Buffalo, State University of New York, Buffalo, NY 14260-1900, USA}
\author{P. A. Bennett}
\affiliation{Department of Physics and Astronomy, Arizona State University, Tempe, AZ 85287-1504, USA}

\begin{abstract}
We explore the problem of tunneling through disorderd nanowires, comprised of a random distribution of metallic grains, by means of a many-body model that captures the essential physics of the system. The random configuration of grains gives rise to a smooth band-like set of states, which mediates current flow through the nanowire. Analytical and  numerical calculations show the characteristic signature of this unusual band-like transport to be a quadratic variation of the current as a function of the applied voltage (i.e. $I\sim V^2$), a variation that is clearly observed in experimental studies of Pt/C composite nanowires.
\end{abstract}
\maketitle

In spite of the vast literature that has appeared over more than fifty years, interest in the problem of conduction in granular systems remains as high as ever, in large part because these systems provide a unique opportunity to study transport under conditions where many-body interactions and disorder both exert a non-perturbative influence on carrier conduction. Recent work on this problem continues to result in predictions of novel phenomena, such as single-electron charge soliton propagation in one-dimensional arrays of metallic grains,\cite{altland2004} and coherent electron tunneling over large distances in systems with large intergranular conductance.\cite{beloborodov2003,efetov2003} Experimental studies have revealed interesting effects, such as intergranular superconductivity\cite{gerber1997} that can lead to global superconductivity, or perfect insulating behavior, depending on the strength of the intergranular Josephson coupling, and spin-dependent tunneling in self-assembled superlattices of magnetic nanoparticles.\cite{black2000}

This paper explores the problem of tunneling through a disordered system, comprised of a large number of metallic grains with a random distribution of sizes, and inter-grain separations, under conditions where hopping via the grains is dominated by single-electron tunneling, and the Coulomb repulsion between neighbouring grains is strong. A theoretical analysis of this system shows that its random configuration smears out any features due to tunneling via the individual grains, and gives rise instead to a smooth band-like set of states through which the current flow is mediated. Analytical calculations (and numerical simulations), reveal the characteristic signature of this unusual band-like transport to be a quadratic variation of the current as a function of the applied voltage (i.e. $I\sim V^2$), without any well-defined threshold for the onset of conduction. Clear evidence for this behavior is found in experimental studies of Pt/C nanowires, which show a quadratic variation of their current for the entire range of voltage studied. We also investigate the temperature dependence of the current in these nanowires, and show that the observed behavior can be accounted for by allowing for a temperature-dependent hopping rate between the grains.

Since the fabrication and basic electrical characterization of the nanowires that we study was described in detail in Ref. \onlinecite{rotkina2003}, for further details we refer the reader to this publication. The nanowires are formed by the technique of electron-beam induced deposition, which yields uniform structures with a typical diameter of $\sim50$ nm. The nanowires are actually comprised of a Pt/C composite, however, with Pt crystallites a few nm in size that are embedded in an amorphous C matrix. To study the electrical properties of the nanowires, we deposit Ti/Au electrodes by electron-beam, and optical, lithography, and make two-terminal measurements of the $I-V$ characteristics of nanowire segments a few $\mu$m in length.\cite{rotkina2003}

\begin{figure}[t]
\begin{center}
\includegraphics[width=6.5cm]{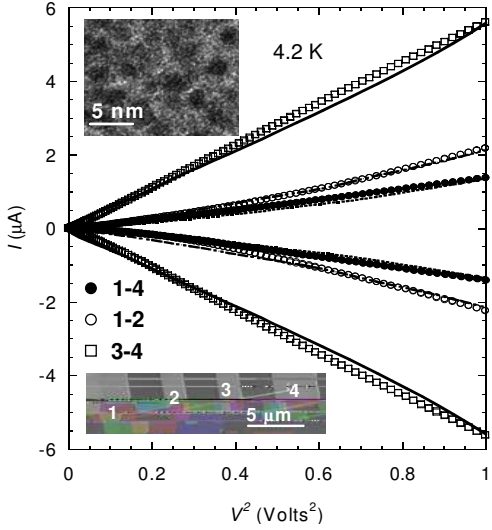}
\end{center}
\caption{Comparison of experimental and theoretical (lines) currents for measurements between contacts 1-2 (rings), 1-4 (dots), and 3-4 (squares), given $t_0\sim0.05$ eV and $\Gamma_0\sim0.3$ eV (see text), at $T=4.2$ K.}
\label{fig-jv}
\end{figure}

\begin{figure}[t]
\begin{center}
\includegraphics[width=6.5cm]{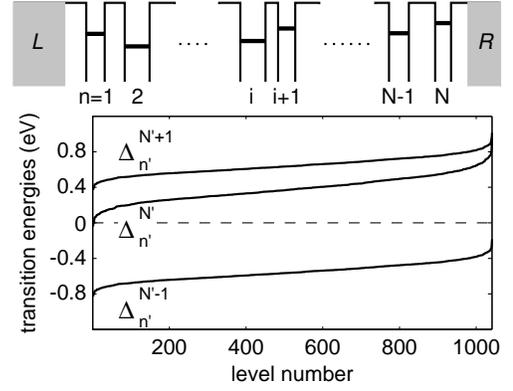}
\end{center}
\caption{Upper panel: One-dimensional array of quantum wells separated by tunnel barriers, connected to the left/right ($L/R$) reservoirs. Lower panel: Transition energies of the QW array in increasing order constituting the band-like set of conductive states.}
\label{fig-array}
\end{figure}

The motivation for our theoretical studies is provided by our recent investigations of the electrical characteristics of Pt/C composite nanowires, realized by the process of Electron-Beam-Induced Deposition (EBID). The fabrication of these nanowires has been described in detail in Ref. \onlinecite{rotkina2003}, where we also reported preliminary measurements of their current-voltage characteristics. While the EBID process yields uniform structures several microns long, with a typical diameter of ~50 nm, as we illustrate in the upper inset to Fig. \ref{fig-jv}, the nanowires are actually comprised of a Pt/C composite, with Pt crystallites a few nm in size that are embedded in an amorphous C matrix. For theoretical modeling of the nanowires, we consider a one-dimensional array of quantum wells (QWs) separated by tunnel barriers that model the insulating C matrix (see Fig. \ref{fig-array}). The QWs are characterised by their widths $\mu^{QW}_i,\ i=1,\dots,N$, where $N$ is the total number of QWs in the array. Due to the finite distance between the grains in the Pt/C wire, the QWs are separated by tunnel barriers of the widths $\mu^{TB}_{ij},\ 1\leq i<j\leq N$. In the experimental structure, the sizes of the Pt grains and the distance between them are random numbers, hence the parameters $\mu^{QW}_i$ and $\mu^{TB}_{ij}$ are generated from Normal distributions $N(\mu^{QW},\sigma^{QW})$ and $N(\mu^{TB},\sigma^{TB})$, respectively, where $\mu^{QW/TB}$ and $\sigma^{QW/TB}$ are the corresponding expectation value and standard deviation. This is motivated by the large amount of Pt grains in the nanowire. Physically, the one-dimensional array of wells describe a single chain of Pt grains in the nano-wire. Since the interactions between the different chains merely extends the set of conductive states in the system, this model provides a reasonable account of the transport properties. 

The largest energy scales in the model are the on-site and intergranular Coulomb interactions (on-site charging energy $U\sim1$ eV and intergranular Coulomb interaction $U_{ij}\gtrsim0.3$ eV\cite{delft2001}). Thus, it is motivated to transfer the model into diagonal form by introducing the many-body operators $\ket{p}\bra{q}$,\cite{hubbard1963} which describe transitions from the state $\ket{q}$ to $\ket{p}$. Diagonalization of the model also gives a complete freedom in the choice of parameters $U_i,\ U_{ij}, t_{ij}$, where $t_{ij}$ is the intergranular hopping. Experimental work has shown a strong gating effect on the $I-V$ characteristics of the nanowire,\cite{rotkina2003} suggesting that the transport through them is of single electron character. Hence it is only necessary to account for the energy separation between states like $\ket{N',n}$ and $\ket{N'\pm1,n'}$, where $N'$ is the equilibrium ground state number of electrons in the nanowire whereas $n,n'$, are state labels. The energy difference between the states $\ket{N',n}$ and $\ket{N'-1,n'}$, denoted by $\Delta_{1nn'}$, corresponds to the energy of the removed electron, whereas the difference between $\ket{N'+1,n}$ and $\ket{N',n'}$, denoted by $\Delta_{2nn'}$, and $\Delta_{3nn'}=\Delta_{1nn'}+U$, corresponds to the energy of the added electron plus the intergranular charging energy. Hence, written in terms of the many-body eigenbasis, the model for the Pt/C nanowire becomes
\[
\Hamil_{Pt/C}=\sum_{m=0,\pm1}\sum_nE_n^{N'+m}\ket{N'+m,n}\bra{N'+m,n},
\]
where $E_n^{N+m}$ is the energy for the corresponding states. In the experimental set-up, the Pt/C nanowire is coupled to external contact reservoirs, here modelled by free electron-like particles. Hence the Hamiltonian for the nanowire and the reservoirs is given by
\begin{eqnarray}
\lefteqn{
\Hamil=\sum_{k\sigma\in L,R}\dote{k}\cdagger{k}\c{k}
	+\Hamil_{Pt/C}
}
\nonumber\\&&\hspace{-0.5cm}
	+\sum_{\stackrel{{\scriptstyle k\sigma,nn'}}{m=0,-1}}
		(v_k\cdagger{k}\ket{N'+m,n'}\bra{N'+m+1,n}+H.c.)
\label{eq-hamiltonian}
\end{eqnarray}
where $\cdagger{k}\ (\c{k})$ creates (annihilates) an electron in the left/right ($L/R$) reservoir at the energy $\leade{k}$, and the hybridisation $v_k$ between states in the reservoirs and the Pt/C nanowire are assumed to be $k$-independent.

In the Hubbard I approximation (HIA),\cite{hubbard1963,franssonPRL2002} the Green function (GF) for the transition $\ket{N',n}\bra{N'\pm,n}$ is given in this model by
\begin{equation}
G_{nn'}(i\omega)=\frac{P_{\gamma nn'}}
	{i\omega-\Delta_{\gamma nn'}-P_{\gamma nn'}\Gamma/2},
\label{eq-GF}
\end{equation}
where $\gamma=1,2,3$, depending on whether an electron is removed or added to the nanowire, whereas $\Gamma=\Gamma^L+\Gamma^R$, and $\Gamma^{L/R}$ is the coupling to the left/right reservoir. The spectral weight $P_{\gamma nn'}$, depending on the bias voltage and the temperature, appears due to the non-trivial anti-commutation relations between the Hubbard operators.\cite{franssonPRL2002} It should be noticed that the GFs for the conducting transitions have to be self-consistently solved for each temperature and bias voltage applied to the system.

By employing the given approximation of the GF to the formula for the current through the system derived in Refs. \onlinecite{jauho1994,franssonPRB2002}, we end up at the expression
\begin{eqnarray}
I&=&\frac{e}{h}\sum_{\gamma=1,2}\sum_{nn'}
	\int\frac{\Gamma^L\Gamma^R}
			{(\omega-\Delta_{\gamma nn'})^2+(P_{\gamma nn'}\Gamma/2)^2}
\nonumber\\&&
	\times P_{\gamma nn'}^2[f_L(\omega)-f_R(\omega)]d\omega,
\label{eq-current}
\end{eqnarray}
where $f_{L/R}(\omega)=f(\omega-\mu_{L/R})$ is the Fermi function at the chemical potential $\mu_{L/R}$ of the left/right reservoir. From this expression for the current, we proceed with analytical calculations by putting the temperature to zero. Thus, the integral in Eq. (\ref{eq-current}) becomes
\begin{eqnarray}
\int_{\mu_R}^{\mu_L}\frac{1}
	{(\omega-\Delta_{\gamma nn'})^2+(P_{\gamma nn'}\Gamma/2)^2}d\omega\sim
\nonumber\\
	\sim\arctan{\frac{\mu_L-\Delta_{\gamma nn'}}{P_{\gamma nn'}\Gamma/2}}
		-\arctan{\frac{\mu_R-\Delta_{\gamma nn'}}{P_{\gamma nn'}\Gamma/2}}.
\label{eq-integral}
\end{eqnarray}

The large number of Pt grains in the nanowire, suggests to replace the sum over $n,n'$ by an integral. For the sake of argument, we assume a rectangular distribution of the transition energies $\Delta_{\gamma nn'}\in[\Delta_\gamma^{\min{}},\Delta_\gamma^{\max{}}]$, thus
\[
\sum_{nn'}\arctan{\frac{\mu-\Delta_{\gamma nn'}}{P_{\gamma nn'}\Gamma/2}}
\rightarrow
\int_{\Delta_\gamma^{\min{}}}^{\Delta_\gamma^{\max{}}}
\arctan{\frac{\mu-\Delta_\gamma}{P_\gamma\Gamma/2}}d\Delta_\gamma,
\]
This integral is analytically solvable and gives
\begin{eqnarray*}
h(x)=\int\arctan{x_\gamma}d\Delta_\gamma=
	-x\arctan{x}+\frac{1}{2}\log{\biggl(1+x^2\biggr)},
\end{eqnarray*}
where $x=(\mu-\Delta)/(P\Gamma/2)$. For $|x|<1$, that is, for states close to the chemical potential, we may Taylor expand this function, giving (letting $\mu_L\rightarrow eV,\ \mu_R\rightarrow0$)
\begin{eqnarray*}
\lefteqn{
h(x_L)-h(x_R)=-\frac{x_L^2}{2}+\frac{x_L^4}{12}
	+\frac{x^2_R}{2}-\frac{x_R^4}{12}+\ldots\approx
}
\\&&
	\approx
	\frac{2\Delta}{P\Gamma}\biggl(eV-\frac{(eV)^2}{2\Delta}\biggr)
		-\frac{4}{3}\biggl(\frac{\Delta}{P\Gamma}\biggr)^3
			\biggl(eV-\frac{3(eV)^2}{2\Delta}\biggr).
\end{eqnarray*}
Here, the behaviour is dominated by the second term since $P\Gamma<1$, showing that the current varies as a quadratic polynomial of the bias voltage. The function in Eq. (\ref{eq-integral}) indicates a staircase like current for each conductive transition. Hence, a smooth character of the total current will arise when the transition energies are smoothly spread out in the energy interval corresponding to the bias voltage applied, with only a few state around the equilibrium chemical potential. This suggests that the transition energy spacing is $\gtrsim0.8$ eV,\cite{transition} for small intergranular hopping. 

In Fig. \ref{fig-jv}, we plot the measured variation of the current in several different nanowire segments (indicated), as a function of the square of the bias voltage, and see that the resulting data indeed fall closely on a straight line, just as suggested by the above analysis. In particular, there is no clear evidence for a threshold for conduction, with the quadratic variation being observed over the entire voltage range.

While the quadratic variation of the current in Fig. \ref{fig-jv} is strongly suggestive of the band-like transport mechanism that we have discussed, we have also performed numerical calculations of the current from Eq. (\ref{eq-current}), and these provide further support for this mechanism. The lines through the experimental data of Fig. \ref{fig-jv} represent the results of these numerical fits, and agree very well with the experimental data. The numerical calculations have been performed with just two free parameters ($t_0$ and $\Gamma_0$). The construction of the one-dimensional array of QWs separated by tunnel barriers, enables an estimation of the hopping parameters $t_{ij}=t_0\exp{(-\mu^{TB}_{ij})}$, where $t_0$ is a parameter that is equal for all tunnel barriers and for all calculations.\cite{hopping} The solution of the resulting eigenvalue system yields the (transition) energies $\Delta_{\gamma nn'}$, see Fig. \ref{fig-array}. Finally, the amplitude of the calculated current in Fig. \ref{fig-jv} has been fitted to the experimental result by means of the coupling $\Gamma=\Gamma_0N^{-1/3}\ (\Gamma^{L/R}=\Gamma/2)$, where $\Gamma_0$ is a fixed parameter accounting for the sum of all chains of Pt grains in the nano-wire.

\begin{figure}[t]
\begin{center}
\includegraphics[width=6.5cm]{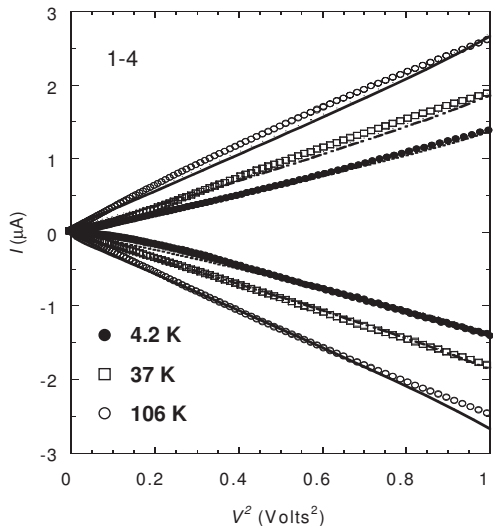}
\end{center}
\caption{Theoretical (lines) vs. experimental (symbols) temperature dependence of the $I-V$ characteristics (measured between contacts 1-4), with $\kappa\sim0.75$ and $\nu\sim0.7$.}
\label{fig-jv_temp}
\end{figure}

The temperature dependence of the current can to good agreement be found partly through the spectral weights, $P_{\gamma nn'}(T)=1-\kappa\exp{(-\nu/[\beta\Delta_{\gamma nn'})]}$, where $\beta^{-1}=k_BT$, from self-consistent calculations of the GFs for finite temperatures, and partly by imposing that the hopping parameter goes like $t_0(T)\sim1+T/T_0$. The latter behaviour is reasonable since the tunnel barriers between the Pt grains in the C matrix become effectively more transparent for higher temperatures. In Fig. \ref{fig-jv_temp}, we display a typical comparison between our experimental and theoretical (solid lines) results for the temperature dependence of the $I-V$ characteristics. For $T>175$ K, leakage currents through the substrate layer are expected to be significant in the experiment, thus we have omitted any calculations for such temperatures. Nevertheless, it can be seen in Fig. \ref{fig-jv_temp} that the theoretical results are in general good agreement with the experimental.

The use of the many-body states in the present model, suggests that we consider the entire Pt/C nanowire as an effective interaction region where electrons can be removed or added via the transitions $\ket{N',n'}\bra{N'\pm1,n}$ at the corresponding energies $\Delta_{\gamma nn'}$. However, the band-like transition energies should not be regarded as energy bands in the usual sense, where an electron at a given energy can propagate freely through the nanowire due to the extension of the energy band throughout the region. Rather, addition/subtraction of an electron to/from the interaction region is followed by a redistribution of the initial state such that the energy for the transition between the initial and final state not exceeds the potential drop over the nanowire given by the bias voltage applied.

In implementing our theoretical model, we have made use of a few approximations in order to obtain the excellent agreement with experiment. The transport properties of the conductive states in the nanowire have been treated using mean-field theory, i.e. within the HIA. This is motivated by the large number of Pt grains that contribute to transport through the nanowires. As transmission electron microscopy has demonstrated, the grains are arranged closely together, although not in thermal contact, and the hopping of electrons between them should tend to smear out the individual character of each grain. Hence, possible charge fluctuations of the individual grains are negligible. Although the transition energies should be renormalized due to strong electron-electron interactions in the grains, such a shift can be included into the mean-field treatement \cite{franssonPRL2002} and will not cause any qualitative change in the nanowire transport properties.

Treating the temperature dependence of the hopping in linear order is motivated by the separation of the energy "bands" ($\gtrsim0.4$ eV), which is much larger than the thermal energy at the temperatures considered here. A more elaborate functional dependence on temperature would possibly alter our results quantitatively, but would not change the general character of the current.

In conclusion, we have explored the problem of tunnelling through disordered nanowires, comprised of a random distribution of metallic grains, by means of a many-body model that captures the essential physics of the system; strong on-site and intergranular electron-electron interactions. Our analysis shows that the random configuration of the grains smears out any features due to tunnelling via the individual grains, and gives rise instead to a smooth band-like set of states through which the current flow is mediated. Analytical calculations (and numerical simulations, involving only a small number of parameters), reveal the characteristic signature of this unusual band-like transport to be a quadratic variation of the current as a function of the applied voltage (i.e. $I\sim V^2$), without any threshold for the onset of conduction. Clear evidence for this behaviour is found in experimental studies of Pt/C nanowires, which show a quadratic variation of their current for the entire range of voltage studied. By imposing a linear dependence of the intergranular hopping strength on temperature, we have been able to obtain a temperature-dependent variation of the nanowire current that is very close to that found in experiment.

J.F. acknowledges helpful comments from O. Eriksson, and support from the Swedish Foundation for Strategic Research (SSF) and G\"oran Gustafsson's foundation.


\begin{thebibliography}{20}
\bibitem{transition} Although the band-like set of transitions depicted in Fig. \ref{fig-array} does not represent all the possible transitions in the nanowire, these are expected to give the major contribution to the conduction. The expansion $\d{i\sigma}=\sum_{pq}\bra{p}\d{i\sigma}\ket{q}\ket{p}\bra{q}$, where $\d{i\sigma}$ annihilates an electron in the $i$th QW, and $\ket{p},\ket{q}$, are eigenstates of the interacting region, contains the transition matrix elements $\bra{p}\d{i\sigma}\ket{q}$ of which a large number have a negligible probability amplitude.
\bibitem{hopping} The exponential decay justifies a simplification of the model to nearest-neighbour interactions in the intergranular hopping. For better accuracy, however, we have used up to the fourth nearest neighbour hopping in our calculations.
\end{thebibliography}
\end{document}